\documentclass[twocolumn,showpacs,preprintnumbers,amsmath,amssymb]{revtex4}

\usepackage{graphicx}
\usepackage{dcolumn}
\usepackage{bm}

\begin{document}

\title{Resonant x-ray emission  spectra of Mn based Heusler alloys.}

\author{M.V. Yablonskikh$^{1,2}$, Yu.M. Yarmoshenko, $^2$ I.V. Solovyev$^{2,3}$, L.
Gridneva$^4$, T. Schmitt $^1$, M. Magnuson$^1$ ,  L.-C. Duda$^1$,
E.Z. Kurmaev$^2$ and J. Nordgren$^1$}

 \affiliation{$^1$Department of Physics, Uppsala University, Angstrom Laboratory, Box 530, S-75121 Uppsala, Sweden}
 \affiliation{$^2$ Institute of Metal Physics, Russian Academy of Sciences-Ural Division, 620219 Yekaterinburg, GSP-170, Russia}
\affiliation{$^3$ Joint Research Center for Atom Technology,
               c/o NAIR, 1-1-4 Higashi, Tsukuba 305-0046, Japan} \affiliation{$^4$
MAXLAB, Lund University, P.O.Box 118, S-22100, Lund, Sweden }

\begin{abstract}
The Mn $L_2,L_3$ x-ray spectra of the Cu$_2$MnAl~and Co$_2$MnZ
(Z=Al,Ga, Sn, Sb) Heusler alloys have been investigated by the
Resonant X-ray Emission Spectroscopy (RXES) using linearly
polarized monochromatic synchrotron radiation for.   The interplay
between the half-metallic character of the Mn 3d electronic
structure in connection with the local magnetic moment $\mu_{Mn}$
and  Mn $2p\rightarrow 3d$ x-ray emission spectra is discussed.

\end{abstract}

\maketitle

\section{Introduction}

The interest to Mn based Heusler alloys \cite{Heusler1} is
connected with the problem of half-metallic ferromagnets (HMF)
\cite{GrootHMFShreder7}. These materials have an energy gap for
one of the spin projections at the Fermi level and represent a
separate class of strong itinerant magnetic substances. The
"classical" alloys of this type are PtMnSb and NiMnSb mostly known
as half-metallic ferromagnets with the lattice structure $C1_b$.
Searching the substances with similar properties it was found the
X$_2$MnZ Heusler alloys with the crystal lattice $L2_1$ where X is
for example Fe, Co, Ni, Pd, Rh and Z is Al, Ga, Si, Sn, Sb, In are
proposed to be also promising for magnetic applications
\cite{GrootHMFShreder7}. Mn-based Heusler alloys are particularly
interesting because of their compatibility with existing
semiconductor technology.  The remarkable feature is a high
magnetic moment (3-4.6 $\mu_B$), that is mainly being formed by Mn
atoms \cite{webster1967}-\cite{IshidaHFM}. So, the electronic
structure of the half-metallic ferromagnets and related materials
allows them to be efficiently applied as a source of
spin-polarized charge carriers \cite{HMFGroot2001}.

 In case of the  X$_2$MnZ the $L2_1$ structure is slightly modified comparing
to $C1_b$, that, however, affects the electronic structure i.e. of
of Mn atoms in a different ways depending on the type of X and Z
elements. The influence to the type of the magnetic ordering has
been predicted and explained mainly because of changes in
intra-atomic distances between the atoms. In particular that
effects the minority-spin electrons to make them lose their
semiconducting character, while the majority-spin electrons keep
to stay metallic \cite{Kublercalc}. Thereby it is important to
understand the interplay between the electronic structure of Mn
and magnetic characteristics of Heusler alloys searching for
relation between the formation of the strong, almost 100\%
spin-polarized structure of the conduction electrons and the
chemical type of the X and Z element. The chance to study topic
has been given by x-ray emission spectroscopy technique using the
ability to extract the element selective information about the
valence structure and detecting the x-ray transitions localized to
the first coordination sphere. We applied the method to study of
the Mn $L_2,L_3$ x-ray emission spectra excited by linearly
polarized monochromatic syncrotron radiation studying
$2p_{1/2}\rightarrow 3d_{3/2}$ and $2p_{3/2}\rightarrow
3d_{3/2},_{5/2}$ photon emission transitions respectively for the
line of alloys Cu$_2$MnAl, Co$_2$MnZ (Z=Al, Ga, Sn, Sb) in order
to investigate the interplay between the local magnetic moment at
Mn site, alterations in the Mn 3d band structure and the
experimental Mn $L_{2,3}$ x-ray emission spectra.


\section{Experimental setup}
The samples investigated are polycrystallines used before at the
series of experiments both by x-ray photoelectron
\cite{Yarmoshenko1,PReB60-6428} and x-ray emission spectroscopy
\cite{OurMCDPR}. The present experiment was performed at the bulk
branch line of beamlne I511 at MAX II (MAX-lab National
Laboratory, Sweden). The x-ray absorption (XA) spectra were
measured by recording the total electron yield (TEY) using the
scanning the photon energy of the incident monochromatized
synchrotron radiation. TEY absorption spectra were recorded by
measuring sample drain current. The XA spectra were normalized to
the photo current from a clean gold mesh introduced into the
synchrotron radiation beam in order to correct for intensity
variations of the incident x-ray beam. The soft x-ray emission
(XE) spectra were recorded with a high-resolution Rowland-mount
grazing-incidence grating spectrometer \cite{ND1} with a
two-dimensional detector. The monochromator energy band pass used
for all XA spectra and for the excitation of the RXE spectra was
approximately 0.15 eV and 0.25 eV respectively. At beam line I511
(MAX II) refocusing optics situated in front of the measurement
chamber and focusing the beam down to a vertical beam size of
below 20 $\mu$ is being used.
  The angle of the incident photon beam was about 5$^\circ$,
while spectrometer had been set perpendicularly in a horizontal
plane. The selected grating and the entrance slit value lead to
the 1.0 eV resolution of the spectra detected. The vacuum was
below $2-3 \cdot 10^{-9}$ Torr.

\section{Results}

\begin{figure}
\includegraphics[scale=0.89]{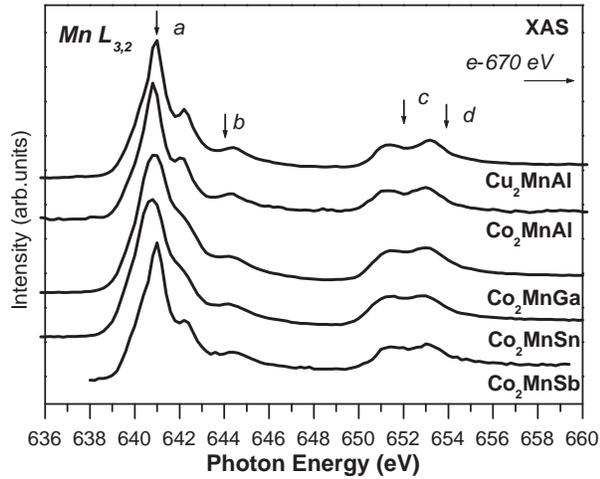}
\caption{XAS spectra. The excitation energies for resonant x-ray
emission measurements are shown by the arrows } \label{figXAS}
\end{figure}

\begin{figure}
\includegraphics[scale=0.89]{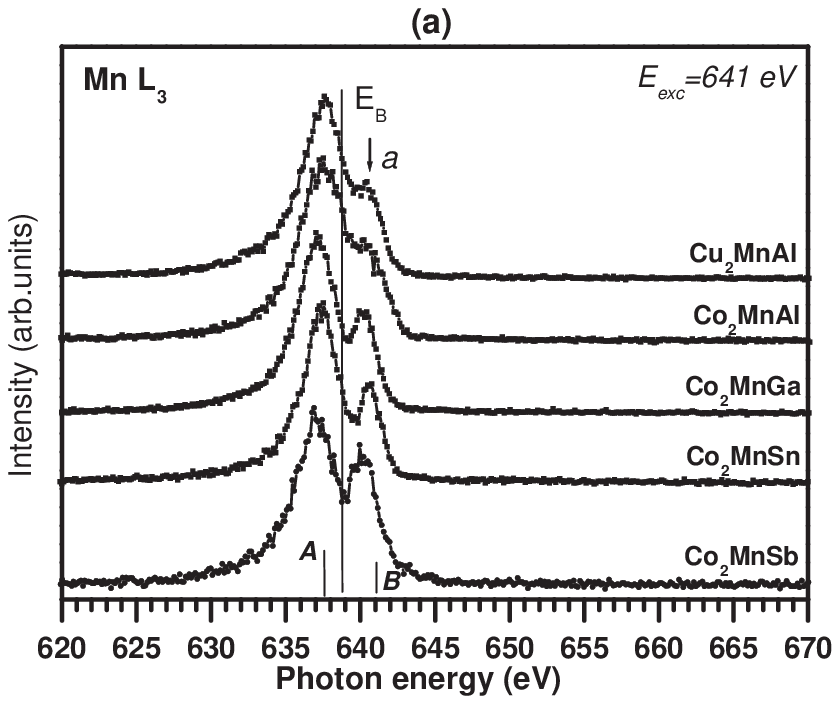}
\includegraphics[scale=0.89]{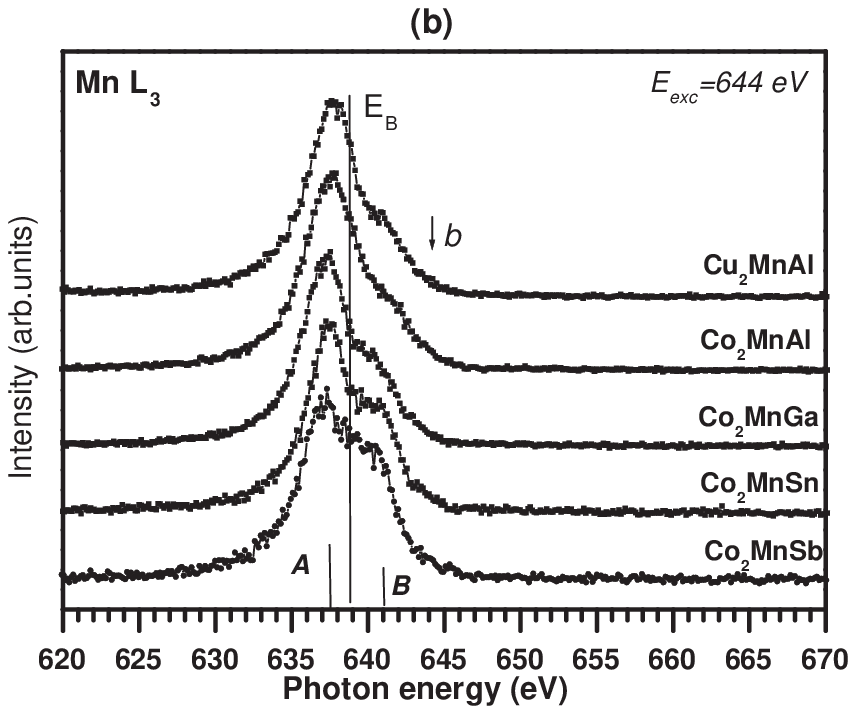}
\caption{Resonant x-ray emission spectra measured by excitation at
the $L_3$ edge (a) and below the $L_2$ threshold. Excitation
energies are indicated by the arrows and marked as $a$ and $b$
relative to x-ray absorption spectrum, Fig.\ref{figXAS}. The
position of the Fermi level E$_B$ is taken from the x-ray
photoemission measurements of the same samples from
\cite{PReB60-6428}, that corresponds to the Mn $2p_{3/2}$ core
level binding energy. The energy position of two peaks forming the
$L_3$ emission line are indicated as \textbf{A} and \textbf{B} }
\label{figL3all}
\end{figure}

  The measured XA spectra are displayed
at the Fig.\ref{figXAS}. They are been looking almost identical to
each other. We found the such spectra are typical of divalent  Mn,
that indicated the metallic character of the chemical bonding.
Presented XA spectra are seems to be almost similar to each other.
Contradictory to that the dependence of the XE spectra for every
excitation energy has some differences from each other. First let
pay attention to spectra at the Fig.\ref{figL3all}(a) excited with
$L_3$ absorbtion edge excitation energy $E_{exc}$=641 eV. The only
Mn $L_3$ x-ray emission line is to be excited at that case. One
can easily distinguish two-peak structure at the region 635-642
eV, where the heights position in the energy scale have been
selected as \textit{A} and \textit{B}.

  For the case of excitation energy $E_{exc}$=644 eV displayed at the
Fig.\ref{figL3all}(b) the two-peak structure is observed also. The
energy position of them is the same like in case
Fig.\ref{figL3all}(a). The intensity of the peak \textit{B}
relative to peak \textit{A} $I(B)/I(A)$ is to be the same for
Cu$_2$MnAl and Co$_2$MnAl but is being increased in line Co$_2$MnZ
from Z=Al to Z=Sb. The dependence is valid both for all cases
presented at the Fig.\ref{figL3all} for both excitation energies
$a$ and $b$ selected below the $L_2$ absorbtion edge.  The elastic
peak with half-width about 0.3 eV is not crossed with the energy
region of the peak $B$ with position at 641 eV in photon energy
scale. Note the $I(B)/I(A)$ ratio is growing up in line of alloys
for the case $a$, Fig.\ref{figL3all}(a), so we still have
$I(B)/I(A)$ dependence in a line Co$_2$MnAl, Co$_2$MnZ from Z=Al
to Z=Sb.

\begin{figure}
\includegraphics[scale=0.89]{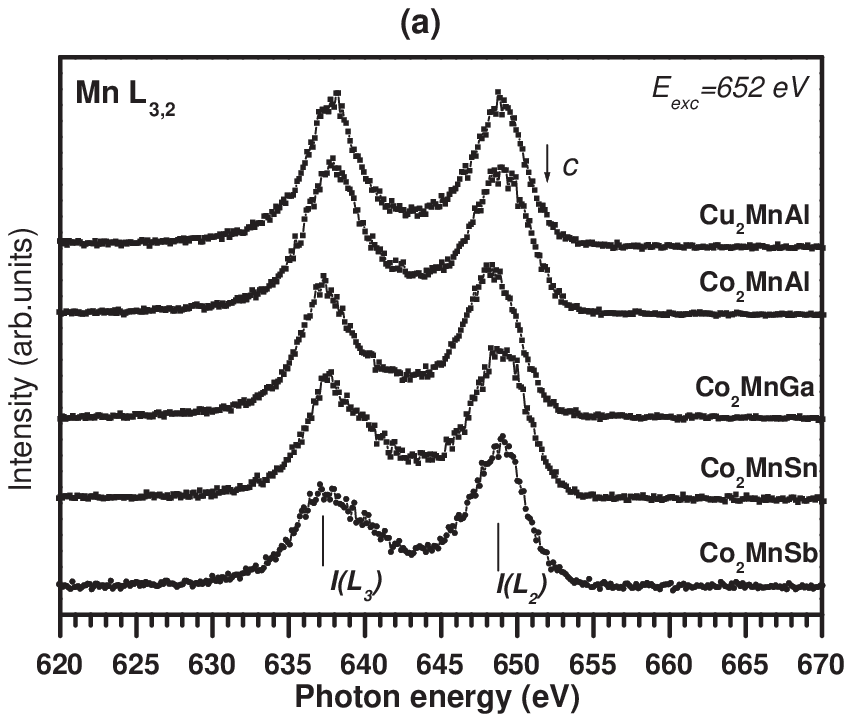}
\includegraphics[scale=0.89]{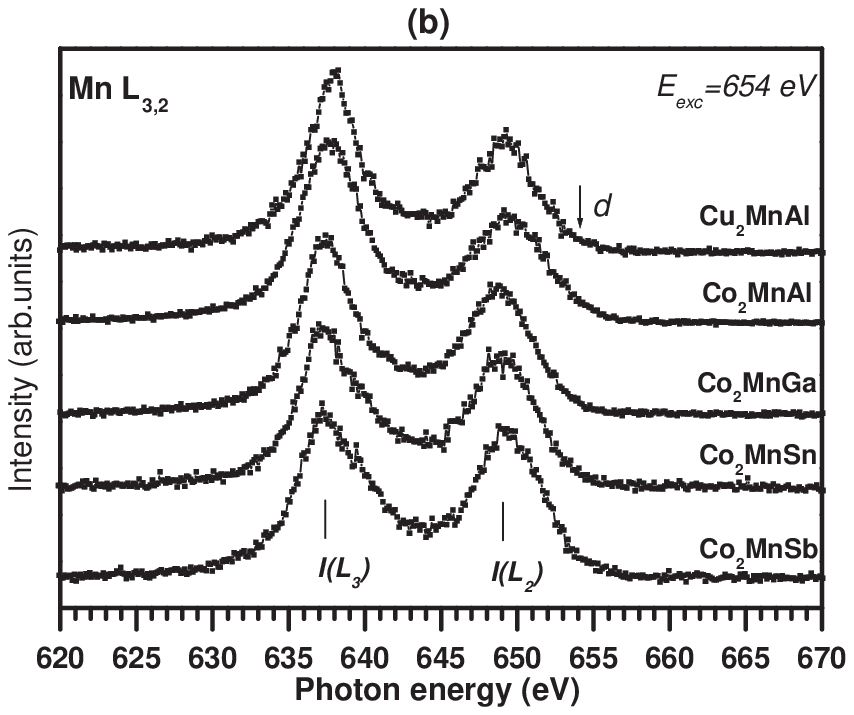}
\includegraphics[scale=0.89]{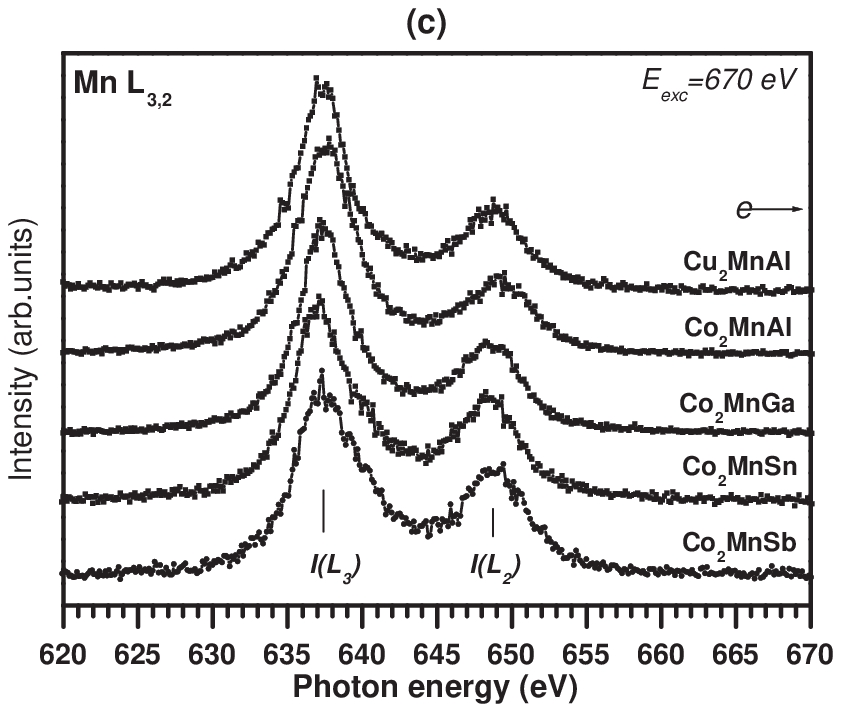}
\caption{Resonant x-ray emission spectra measured by excitation at
the $L_2$ edge (a) and close to the $L_2$ threshold (b) toward
high energy. The spectrum (c) is excited with an energy selected
far from the $L_{2,3}$ absorbtion edges. Excitation energies are
indicated by the arrows and marked as $c$,$d$ and $e$ relative to
x-ray absorption spectrum, Fig.\ref{figXAS}. The energy position
of the $L_3$ and $L_2$ lines are indicated as $I(L_3)$ and
$I(L_2)$.}
 \label{figL2all}
\end{figure}

Next we will pay attention to the spectra excited with energies
near the Mn $L_2$ absorption edge, Fig.\ref{figL2all}(a),(b). The
non-resonant spectra excited with $E_{exc}$=670 eV are shown at
the Fig.\ref{figL2all}(c). The dependence of the $I(L_2)/I(L_3)$
ratio in line of alloys Cu$_2$MnAl, Co$_2$MnAl and to Co$_2$MnSb
is observed, which is similar for $I(B)/I(A)$ ratio. The two-peak
structure of the $L_3$ line can not be distinguished because of
the excitation energy selected far enough from the $L_3$
absorbtion edge.   The dependencies of the $I(B)/I(A)$ and
$I(L_2)/I(L_3)$ from the alloy type are summarized at the
Fig.\ref{figStatistica}.

\begin{figure}
\includegraphics[scale=0.79]{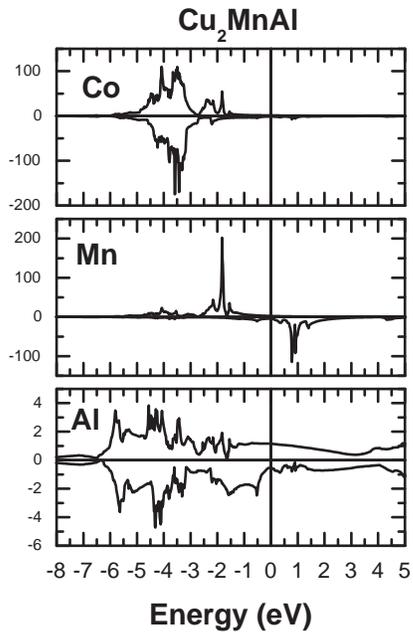}
\includegraphics[scale=0.79]{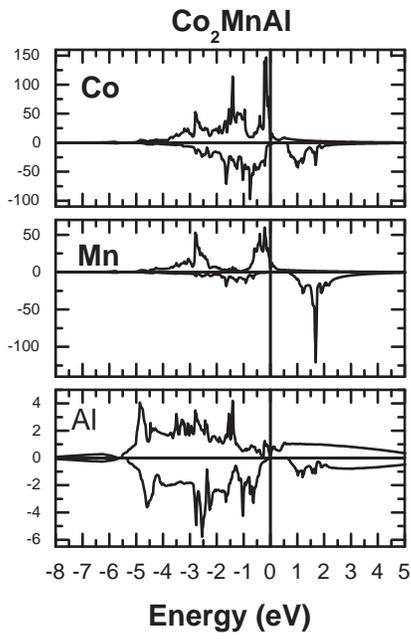}
\caption{Site- and spin- projected state densities of $d$ and $p$
electrons of the Me,Mn and Al } \label{figAlAl}
\end{figure}

\begin{figure}
\includegraphics[scale=0.79]{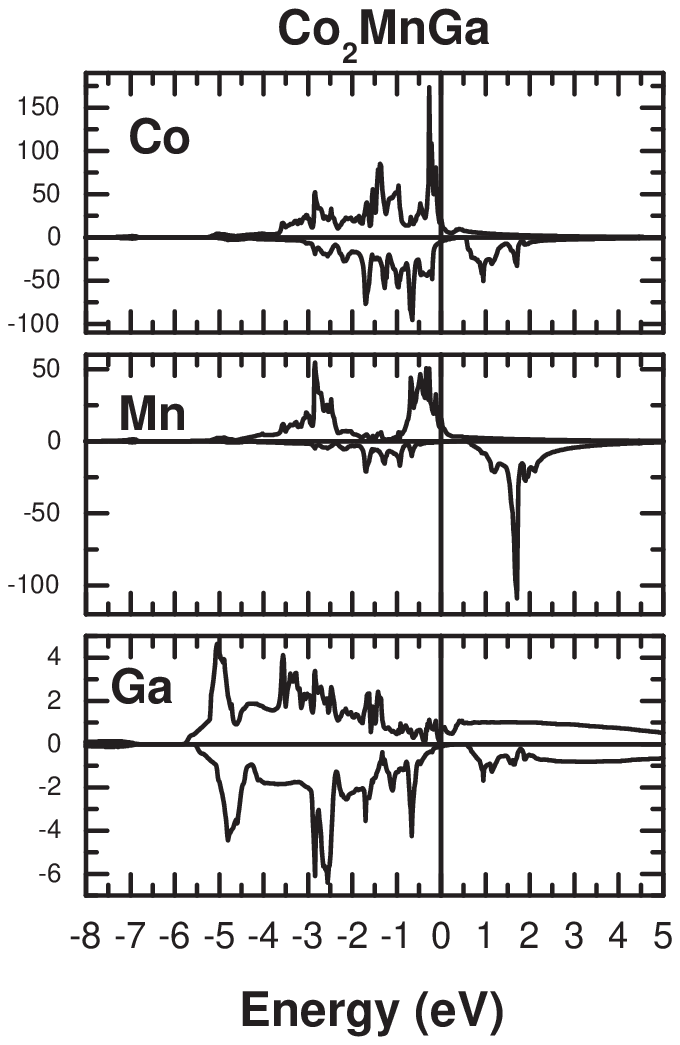}
\includegraphics[scale=0.79]{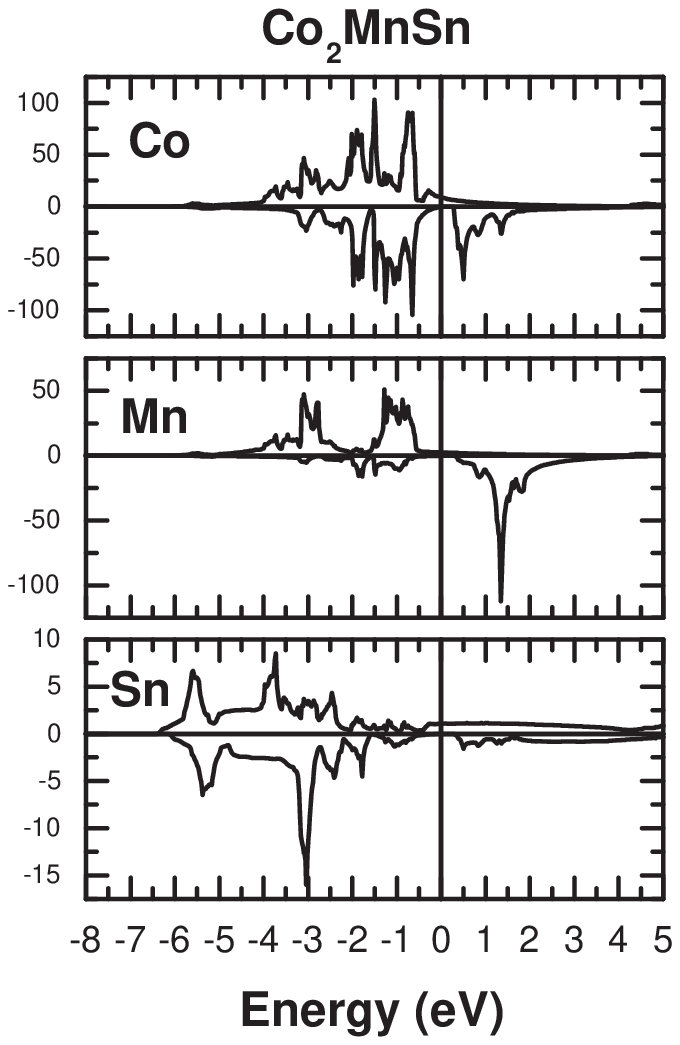}
\includegraphics[scale=0.79]{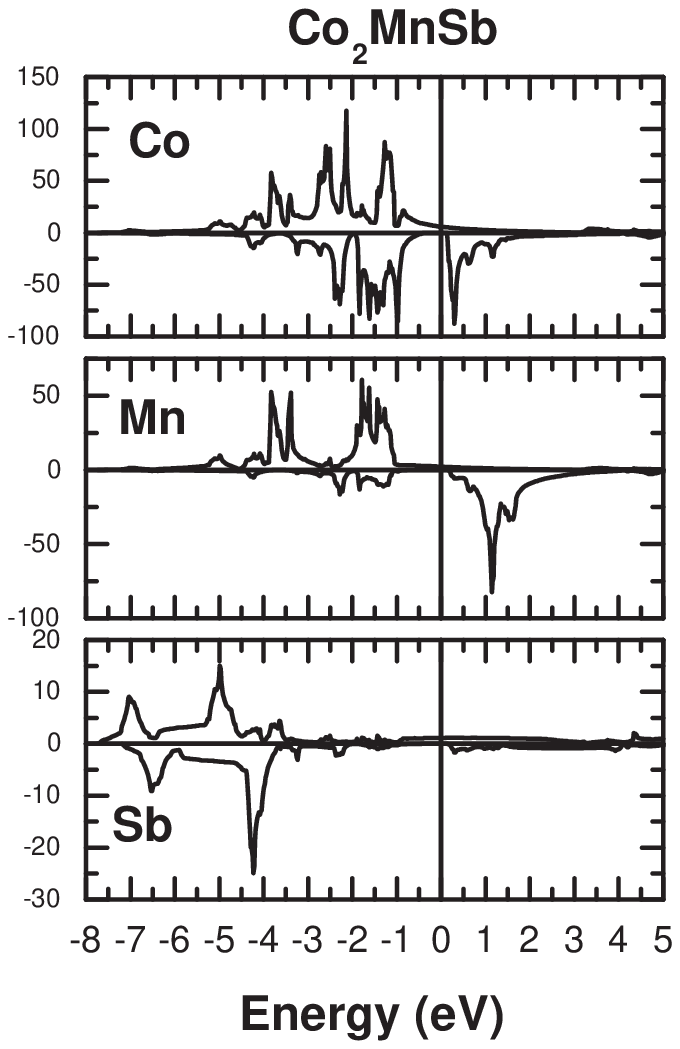}
\caption{Site- and spin- projected state densities of $d$ and $p$
electrons of the $Me$, Mn and $Z$, where Z is Ga,Sn,Sb
respectively  } \label{figGaSnSb}
\end{figure}

The electronic structure calculations in the local-spin-density
approximation have been performed using the linear muffin-tin
orbital method.\cite{LMTO} The atomic spheres radii for X, Mn and
Z were choosen from the charge neutrality condition inside the
spheres, and using the experimental lattice parameters. The
Brillouin zone integration was performed using the linear
tetrahedron method and 172 nonequivalent ${\bf k}$-points
(corresponding to the 14:14:14 devisions of the reciprocal lattice
vectors for the face-centered cubic structure). State densities
are given at Fig.\ref{figAlAl} for X$_2$MnAl (X=Co and Cu) and at
Fig.\ref{figGaSnSb} for Co$_2$MnZ (Z=Ga, Sn, Sb). Here $\uparrow$
designates the majority-spin electrons and $\downarrow$ the
minority-spin electrons. The minority-spin state densities at the
Fermi energy for Mn in Co$_2$MnZ are nearly vanished. Estimated
magnetic moments (Table.\ref{tab:table1}) are in a good agreement
with experimental ones.

\section{Discussion}
First the general terms using for the spectroscopy data will be
briefly mentioned. At the second, the attention will be given to
the calculations and conclusions related. Finally we will finish
with the explanation of the experimental results, considering the
task to not exactly identify the mechanism of the forming Mn
$L_2,_3$ spectra in Heusler alloys, but to find out and describe
the correlation between magnetic properties of the materials,
electronic structure of the Mn 3d in alloys and respective x-ray
emission spectra.

 The shape of the $L_3$ emission line $2p_{3/2}\rightarrow 3d_{5/2},_{3/2}$
reflects 3d DOS only indirectly, first because of the general
nature of the process of resonant/non-resonant photon scattering.
In order to understand the complication of the principal ability
of the spectra formation exactly for Mn $ L_{3,2}$ one can look
into recent work by F. Borgatti \textit{et.al.} \cite{Borgatti}.
 In order to avoid the point of discussion from the general
discussion  of the elastic and inelastic x-ray scattering in soft
x-ray region that is supposed to be the separate branch of
studies, we will focus mostly at the relative changes at x-ray
spectra.

Relating to the experience of x-ray emission studies already been
done for the Mn $L_2,3$ of the Co$_2$MnSb, NiMnSb using the
magnetic circular dichroism the terms of "normal" emission and
"re-emission" has to be only briefly introduced. By "normal"
emission one can understand the $2p\rightarrow3d$ transition
created by the radiative relaxation of the 3d electron from the
\textit{occupied} part of DOS to the 2p core hole. By
"re-emission" the same radiative relaxation of the 3d electron
excited to the \textit{unoccupied} part of DOS is defined. Looking
at the Figs.\ref{figL3all} one can distinguish these impacts to
the Mn $L_3$ spectra: the peak \textit{A} corresponds to "normal"
emission impact and the peak \textit{B} corresponds to the
"re-emission" of the Mn 3d electron from \textit{unoccupied} to
the Mn $2p_{3/2}$ core hole.

 At the Fig.\ref{figAlAl} calculations of the Cu$_2$MnAl
and Co$_2$MnAl are presented. The Me 3d majority-spin electrons in
case of Co are more delocalized then in Cu case. The value of the
Mn 3d density of states  $N(E_F)_\downarrow$ for Mn minority-spin
projection is significantly lower in case of Co$_2$MnAl comparing
to Cu$_2$MnAl, see~Table\ref{tab:table1}.   In line of the
Co$_2$MnZ where Z is being changed from Z=Al to Z=Sb, the the $Co$
$3d$ and Mn $3d$ DOS structures are similar to each other for the
current alloy, see Fig.\ref{figGaSnSb}.  The value of the
$N(E_F)_\downarrow$ is diminished in line Z=Al, Ga, Sn, Sb, that
means the Mn 3d DOS becomes close to the half-metallic state
following the line mentioned.

It is not possible to pass over the systematic calculations
performed for Fe$_2$MnZ ($Z=Al,Si,P$)\cite{Fe2MnZCalc}. The option
discovered for this line tells about the ability to make Mn 3d DOS
structure half-metallic, changing the $Z$ element. That shows
principal ability to reach the half-metallic state of the DOS
changing the chemical type of the third element in X$_2$MnZ. The
similar role of Z atom was also demonstrated experimentally
\cite{webster33,webster81}. That is in a good agreement with
results performed for the Cu$_2$MnAl and Co$_2$MnZ (Z=Al, Ga, Sn)
\cite{Kublercalc} and with presented calculations.

\begin{figure}
\includegraphics[scale=0.5]{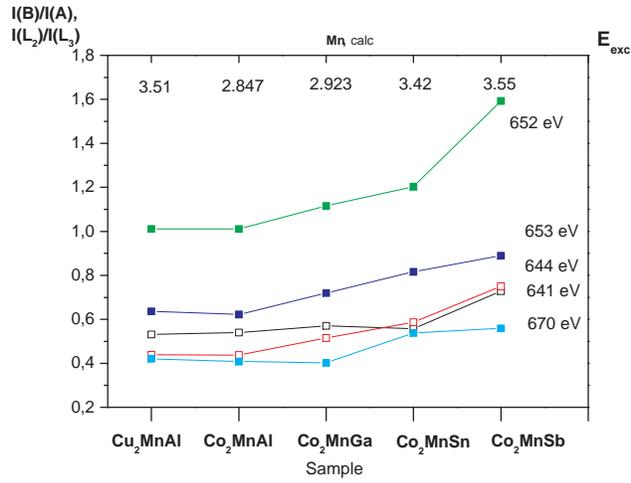}
\caption{Intensity ratio dependence: $I(B)/I(A)$ for  641 - 644 eV
and $I(L_2)/I(L_3)$ for 652-670 eV excitation energies $E_{exc}$,
eV. Excitation energies are at the left grid. Calculate values of
$\mu_{Mn}$ are shown at the top. } \label{figStatistica}
\end{figure}

Now we will focus on the explanation of the dependencies presented
at the Fig.\ref{figStatistica} for the Co- Heusler alloys. Let us
pay attention to the Fig.\ref{figStatistica} and data from the
Table \ref{tab:table1}. In line Co$_2$MnZ (Z=Al,Ga, Sn, Sb) where
the dependence of the $I(B)/I(A)$ intensity ratio  for the each
excitation energy at and above the $L_3$ Mn absorbtion edge
exists. The re-emission peak intensity grows with the decrease of
the  $N(E_F)_\downarrow$.  The conclusion coming proposes the
re-emission as an indicator of the closeness of the Mn 3d band to
the half-metallic state.

The $I(L_2)/I(L_3)$ ratio is also growing in the same line of
Co-alloys with Z=Al,Ga, Sn, Sb comparing to that one in pure Mn.
We have no such a transparent explanation of that. The promising
way seems to consider Coster-Kronig impact in the redistribution
of the $L_3$, $L_2$ intensities. The Coster-Kronig is supposed to
be suppressed for the Mn in Heusler alloys that is found to be
opposite for the case of pure Mn \cite{PReB60-6428}, but we cannot
estimate the actual impact of the process in current case.

 One can not also exclude the possibility of the re-emission
to be involved in the formation of Mn $L_2$ x-ray spectra, that
here is the radiative relaxation of the 3d electron from
unoccupied part of Mn 3d DOS to the Mn $2p_{1/2}$ core hole. The
conclusion one can make in this case is the connection between the
growth of the $I(L_2)/I(L_3)$ ratio and the degression of the
$N(E_F)_\downarrow$ value for the Z changing from Al to Sb in line
of Co- alloys. We hope that experimental fact  to be explained
lately and be accepted as a observed.

 The numerical values at the graph at Fig.\ref{figStatistica} make
us believe in the existing correlation of the $I(B)/I(A)$ and the
$I(L_2)/I(L_3)$ ratios with the value of the magnetic moment
$\mu_{Mn}$ in line of Co$_2$MnZ, where Z is changing in order Al,
Ga, Sn, Sb.

 The situation becomes more complicated with the consideration
of the Mn $L_{2,3}$ XE-spectra of Cu$_2$MnAl in connection with
data for Co- Heusler alloys. The numerical values of the
$I(B)/I(A)$ and the $I(L_2)/I(L_3)$ for X$_2$MnAl alloys are
almost the same.   Due to the suggested mechanism of the $L_3$
spectra formation $I(B)/I(A)$ should be much smaller for
Cu$_2$MnAl, because of the $N(E_F)_\downarrow$ is much higher for
Cu$_2$MnAl then for Co$_2$MnAl.  There is also no correlation
between the intensity ratios  $I(B)/I(A)$, $I(L_2)/I(L_3)$ and
$\mu_{Mn}$ for the Cu$_2$MnAl, Co$_2$MnAl group having different
both $\mu_{Mn}$ and $a$ lattice parameter,
Fig.\ref{figStatistica}, Table.\ref{tab:table1}. The same
$\mu_{Mn}$ and $a$ values had been measured for the Cu$_2$MnAl,
Co$_2$MnSb, but the $I(B)/I(A)$ and $I(L_2)/I(L_3)$ parameters are
different for alloys.

  According to that  it is difficult to declare the transparent
correlation between the dependence of the $I(L_2)/I(L_3)$,
$I(B)/I(A)$ with both  values of the $\mu_{Mn}$ and the
$N(E_F)_\downarrow$ values for the X$_2$MnZ where Z is constant
element.  The strong difference  of Me 3d and Mn 3d band in case
of Co- alloys and Cu- alloy could be a reason for that. The
correlation is to be valid  in the case X$_2$MnZ alloys, where Z
is changing with the constant metal X.

\section{Conclusion}

There are several leads to be expressed. It was found the relative
intensity of re-emission channel to be intensive and dependent on
the value of the density of states at the Fermi level for the
minority spin states in the half-metallic alloys of X$_2$MnZ type,
where X metal is to be constant. Both the growth of the
re-emission impact in forming Mn $L_3$ spectrum and
$I(L_2)/I(L_3)$ ratio are connected with the decrease of the
$N(E_F)_{\downarrow}$ value.

 Having an example of the Cu$_2$MnAl and Co$_2$MnAl one can suggest that
such an simple experimental approach of the estimation the gap
value to be not functional in case of the replacement of the $X$ -
metal  in $X$$_2$MnZ Heusler alloys.  That is connected with a
strong hybridization impact from Me $3d \uparrow$ and Mn $3d
\uparrow$  states.

The result being shown could be used as a basis to continue the
experiments with linearly or circularly polarized synchrotron
radiation defining not only energy distribution of the scattered
photons, but the overall intensity for the different excitation
energies.

\section{Acknowledgment}
Support from the Royal Swedish Academy of Sciences for cooperation
between Sweden and the former Soviet Union is gratefully
acknowledged. We would like to thank M. Katsnelson for the
fruitful discussion and MAX-Lab National Laboratory staff for
experimental support.

\newpage
\begin{table*}
\caption{\label{tab:table1} Structural, magnetic parameters of
Heusler alloys through experimental data of
\cite{SilvaTable},\cite{bushovShreder46},\cite{GrootHMFShreder7},\cite{OxleyHeusler},\cite{FujiiTable},\cite{IshidaHFM}.
Calculated values of the magnetic moments are indexed with $
calc$. The Fermi-level state density of  Mn 3d minority-spin down
states is taken from the calculated figures of the DOS
respectively, eV $^{-1}$ }
\begin{ruledtabular}
\begin{tabular}{ccccccccc}
alloy  & struct. & $a$ , $nm$ & $\mu_{\mbox{Mn}}$ , /
$\mu_{\mbox{Mn calc}}$, $\mu_B$ & $\mu_X$ / $\mu_X{\mbox{ calc}}$
, $\mu_B$ &  $E_f$ Mn 3d$\downarrow$, a.u.\\
 \hline
Cu$_2$MnAl & $L2_1$ & 0.5949 & 3.49/3.51& <0.1/-0.005 & -4.6 &\\
Co$_2$MnAl & $B2/L2_1$& 0.5756 &3.01/2.847 & 0.50/0.684  &-0.23&
\\ Co$_2$MnGa & $L2_1$        & 0.5770 & 3.01/2.923 & 0.52/0.687 &-0.20&
\\ Co$_2$MnSn & $L2_1$        & 0.6000 & 3.58/3.420 & 0.75/0.864 & -0.10& \\
Co$_2$MnSb  & $L2_1$  & 0.5929 & 3.58/3.552 & 0.75/0.983 & 0&
\\

\end{tabular}
\end{ruledtabular}
\end{table*}

\newpage

\bibliographystyle{Prsty}
\bibliography{all}
\end{document}